# 5G Network Management, Orchestration, and Architecture: A Practical Study of the MonB5G project

Hisham A. Kholidy, Mohammed Abuzamak,
State University of New York (SUNY) Polytechnic Institute, Network and Computer Security Department, Utica, NY USA
hisham.khoidy@sunypoly.edu, abuzamm@sunypoly.edu

***Abstract-*** *The cellular device explosion in the past few decades has created many different opportunities for development for future generations. The 5G network offers a greater speed in the transmissions, a lower latency, and therefore greater capacity for remote execution. The benefits of AI for 5G network slicing orchestration and management will be discussed in this survey paper. We will study these topics in light of the EU-funded MonB5G project that works towards providing zero-touch management and orchestration in the support of network slicing at massive scales for 5G LTE and beyond.*

***Key Terms-*** 5G, Network, Management, Orchestration, Architecture, Network Slices, Mobile Core, MonB5G, IDMO, O-RAN, eCPRI.

## 1. INTRODUCTION

The expansion of technology in the past two to three decades has created a new culture that will be the new Norm for a very long time. Cellular devices are one of the main technologies that have thrived in the past decade leaving most people with a phone in their pocket. This is the reason why the development of the new 5G network is extremely important for future generations. In this survey, we will establish the main components of a 5G network architecture 5G network management, and the practicality behind the 5G network orchestration.

## 2. Basic 5G Architecture

The basics behind the 5G architecture is that it helps connect devices wirelessly. These devices currently consist of cellular devices and or tablets but in future reference, it will include many different things for example a vehicle or medical technology.

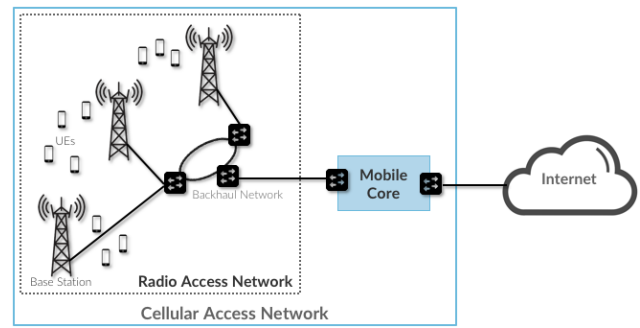

***Fig 1.*** *This figure shows the basics behind the 5G architecture.*

To help break down is 5G Network architecture, and uses two primary aspects as seen in the figure. The first aspect that it uses the Mobile Core which has many different capabilities. Mobile core helps Trace the user to assure that they don't have
a disconnect in their service and are getting billed and charged correctly. It also provides IP correspondence for cellular data. The second aspect is the Radio Access Network (RAN). The radio Access Network helps
keep track of the radio range and ensures that the end-user gets the caliber of quality they are promised.

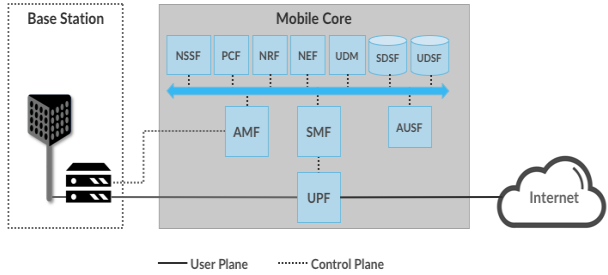

**Fig 2.** Mobile Core

### 2.1. Mobile Core

The 5G mobile core can be broken down into three different main groups. The first part is the Control Plane which helps managers to the 5G system and connects it with also access authorization and Authentication which is also a main aspect of the 5G Network. It also has an authentication server which helps smooth out the interaction between the server and the end user. Also, what differentiates this part is that it has a direct counterpart. The second part which is in the control plane as well doesn't have a direct counterpart. The second part of the control plane is a service that helps create a flow of data that is structured. It also helps discover services that are available. The last part cuz in the user plane is what helps Route traffic between the internet and the RAN. This last part that runs on the user plane also has a responsibility for any lawful issues or developments that are occurring within the system.

### 2.2. Radio Access Network (RAN)

The Radio Access Network (RAN) is the mediator between the end-user and the service provider. It helps navigate throughout the network and makes sure that the end-user receives the quality content that they deserve.

### 3. Concerns Regarding Orchestration and Management

There is a common misconception that network slice management requires a single network domain but that isn't the case because it is not classical network management that requires a single network domain instead requires multiple network domains. Currently, there are MANO solutions that realistically only focus on a more centralized perspective which if there were multiple technical or administrative domains that would produce a lot of scalability problems, especially with those that are visualized in the network slicing conditions. When this case has fully deployed the communication and interaction between the distributed network entities and orchestration network entities we'll have an extreme amount of traffic overload and when that occurs it will stop the execution of standard polling-based monitoring. There has to be a deployment of an extremely lightweight and structured operation that involves closed-loop feedback to be able to have assurance but there is up-to-date monitoring while the resource allocation is happening. Currently, the MANO framework is not able to manage the effects that could occur to the resources that are across many network centers and domains. The orchestration is required to be able to title when there are different networking functionalities that are operating and all of that would be based on the location monitoring of any exploit information.

### 3.1. Network Slicing

A network slice can be observed as the Configuration of soap slices that have affiliations with many different technological domains. To be able to go over this hurdle each technological domain has to have a designated management agent which has to be closer to the resources to be orchestrated. When that happens it will help enable faster reactions and detection of problems. This is where hierarchical orchestration is introduced. A hierarchical orchestration establishes a function that helps support the different amounts of levels of centralization depending on the current bases of operation status and then is able to analyze and make decisions. Lower levels create a limitation over the monitoring overhead which reduces the reaction time of the management decisions. The RAN network slice subnet instance (NSSI) and the CN NSSI are two subsets that can be created from a network slice. The network slicing and services are independent, and a slice can support multiple services at once. It is particularly versatile because the network slicing could still exist after the service is finished. The figure below displays a diagram of Network Slicing.

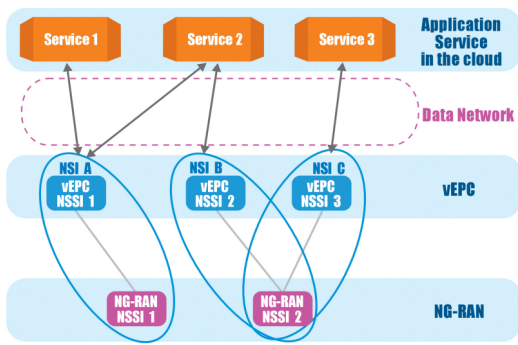

*Fig 3.* Application Service in the Cloud

### 3.2. Pros of AI for Network Slices

5G helps initiate the usage of virtualization technology that way it is able to provide personalized services that are capable of providing the needs of the end-user and the end-user can be anyone which makes it a lot more expendable. The slices have a set of virtual network functions that execute certain things that the slice needs to be able to have a useful service. In so the cloud-based links of the network slice are connected to the literal links but when it comes to virtual network functions they are connected to physical nonce within the mean framework. The utilization of machine learning and artificial intelligence have been adopted throughout the world. It is said to be the future and will affect many different things like functions and the ability to be able to develop a future mobile network. ML and AI already are able to execute task-level issues that copy humans and because of that, it leaves a huge future for both because it could be developed a lot more and be established at a higher level in the near future. The operational keys of mobile networks and network management both have something in common which is AI implemented in both. Tasks that previously took an extremely slow time for humans to evaluate and execute upon snow being processed by an AI an ML unaided by a human can perform it. AI is the apparent running row in the future to be fully automated and do minor tasks that would take a human a long time to do, remotely, and at a much faster rate—being able to statistically manage the slices By making sure that resources are being allocated to match the end-users demand with the addition of being able to do slice admission control.

### 3.3. Virtual Evolved Packet Core (vEPC)

Numerous virtual network functions are created using LTE EPC functionalities (VNFs). The capacity to swiftly deploy service environments and save construction costs is provided by virtualization. As seen in the figure below several vEPC components are addressed below. The roles of these parts are contrasted with those of LTE EPCs.

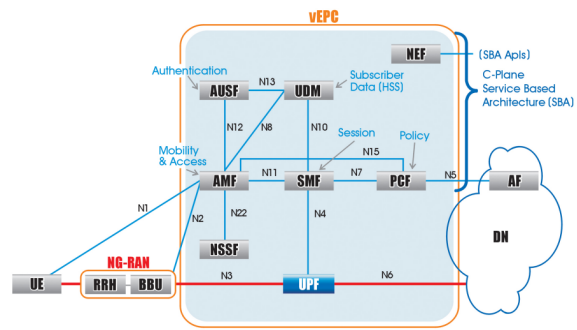

*Fig 4.* vEPC

Numerous NFVs have virtualized NG-RAN and vEPC. The NFV system is more expansive and adaptable than the telecom operators' prior business strategy. Combining the VNF into a network function chain, which determines how to transfer packet flows from this VNF to another, is required for the development of an NFV system. The complex design of 5G networks necessitates an effective management solution. The goal of NFV MANO systems is to manage running services, process network resources, lower deployment costs, and simplify the implementation of new services.

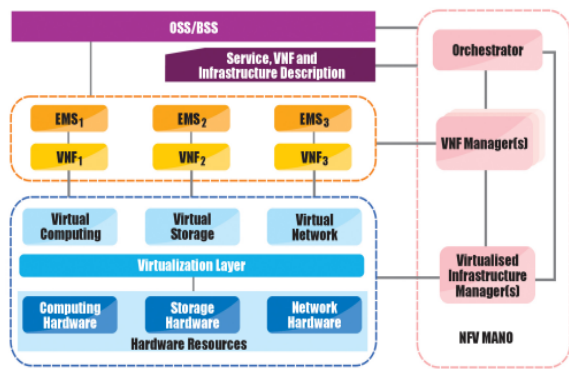

*Fig 5.* VNF and Infrastructure Description

### 3.4. Security Identification

Denial of service attacks (DoS) is one of the key malicious activities that help infiltrate a system. There are two types of Denial of Service attacks (DoS) that we should be aware of and have prevention techniques against; the first is flooding and the second is semantic. Additionally, the performance analysis in the following section demonstrates that SBCA can handle a lot more queries than traditional centralized systems. The second is DoS Flooding Attack Prevention by decentralizing the authentication effort from the AUSF/ARPF to all gNBs, SBCA reduces the effects of legitimate requests flooding. The total computing power of gNBs is growing along with the number of gNB deployments, making it more difficult for adversaries to overwhelm and bring down the entire 5G network. However, the performance analysis in the following section demonstrates that SBCA can handle a lot more queries than traditional centralized systems. As a result, listeners are unable to recognize a particular device or user error messages to track down a device. With SBCA, the activation key is generated using the public ECDH keys that are generated at random. As a result, attackers will not be able to recover the session key even if all temporary keys are taken in the future. If the session key is compromised that won't affect its confidentiality.

### 3.5. Assessment and Results of the Performance

We modeled all centralized protocols to be the M-M-1 exhibiting model and our SBCA
to be the M-M-c waiting in order to show the latency of protocols when there are many legitimate requests. For the sake of simplicity, we assume that the locations of UEs inside the 1 km2 area and the arrivals of requests follow a Poisson distribution. Expect that the centralized protocols' server-side operations will be 80% quicker.
It is expected that gNBs have an ISD of 200m. Every server must therefore respond to queries from 1km^2/0.1^2 = 31.8 ≈ 32 gNBs. On a machine with an Intel Core
i5-3210M with a 2.5 GHz CPU and 16GB of RAM, all simulations are run. The simulated code is written using Charm Crypto with Python 3.7 and adheres to NIST Suggestion to utilize 256-bit similar key strength across all simulations. The empirical computational overheads of various approaches are shown below as the last point. It demonstrates that SBCA has inevitably added 2.84 ms of extra computational overhead above that of 5G-AKA.

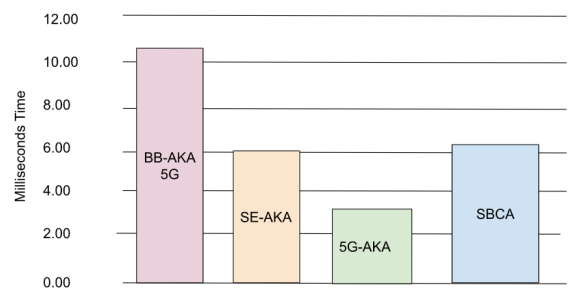

*Fig 6.* Calculations Delays

In TS 22.26 for wireless connections and 1Gbps optical fiber for wired connections inside 5GC in terms of communication overheads. 50 meters is all that separates two operational entities within the 5GC. While the distance between gNB and the closest functional entity is estimated to be 1 km. We further assume that the blockchain nodes' databases can be synchronized.

Synchronize their databases concurrently without noticeably slowing down transmission. The graph below shows how the totals of the transmitting and dispersion delays were calculated. It demonstrates that from 160-bit to 384-bit key lengths, SBCA achieves the lowest overall communication latency. Although the overall latency of SBCA is 3.43 us in the 512-bit higher than 5G-AKA.

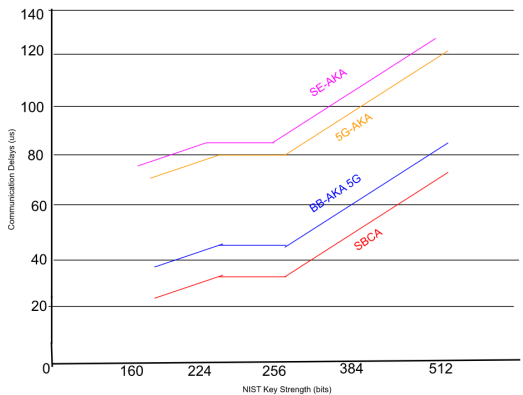

*Fig 7. Communication Delays*

DoS attacks are now more likely due to the fast development of mobile penetration in the 5G era. To safeguard 5G networks from denial-of-service attacks and other severe network risks, a secure blockchain-based 5G authentication and key agreement protocol are recommended. The performance evaluation also showed that SBCA has enhanced the 5G authentication security with the right number of overheads, supporting the security study's conclusion that SBCA is safe and resistant to a variety of network assaults.

**4. MonB5G Architecture**

The MonB5G architecture Executes upon the management system decomposition. The MonB5G expands on understanding the problems that occur with the distribution of orchestration and management for network spacing. The architecture also focuses on the life cycle management making sure that all the resources and how's it going man are agnostic to slices. This way it provides extreme amounts of benefits because it helps isolate the management planes of the slices. The design behind the architecture is a feedback loop that controls different loops with different timescales.

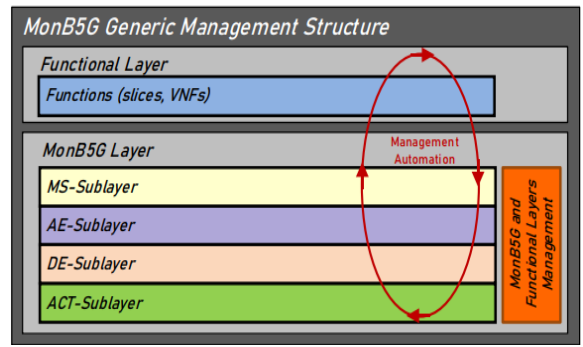

*Fig 7. Slice Structure of MonB5G*

The first layer of the MonB5G framework is orchestration and management which has the framework's fundamental duties and is accountable to the segments of slices. The second layer is the infrastructure layer which relies on the construction and management layer because it enables it to be able to use its layer to full use and also relies on its providers and infrastructure. The last layer is the business layer which is business organizations that utilize the framework and manage the slice services. The MonB5G framework also has static components which help show the capabilities of the MonB5G framework and is also used by the MonB5G system operators. In the SBCA, UE calculates a new message using its chameleon hash trapdoor. Given that it is without knowing the trapdoor, it is practically difficult to establish a hash function collision, so UE effectively proves to the gNB that it is the legitimate originator of the authentication request. After the procedure has been finished, both sides can mutually authenticate each other.

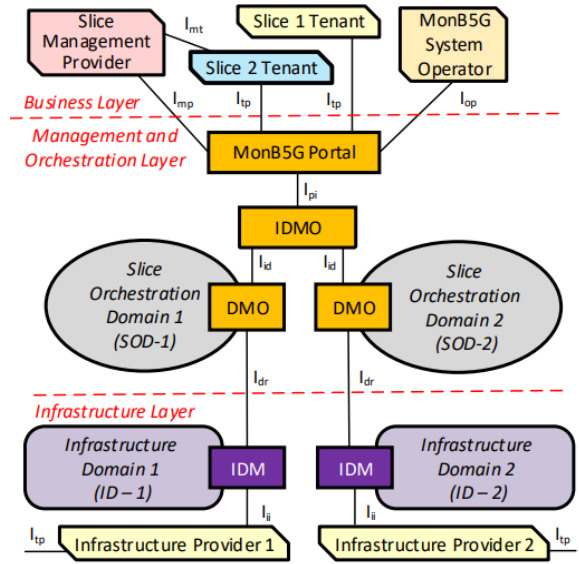

***Fig 8***. *Static & Business Components of The MonB5G Architect*

The above figure shows the static components of the MonB5G mixed in with the business layer. The MonB5G Portal houses the slice management providers, infrastructure providers, and slice tenants.
The MonB5G Portal is also used to help provide billing and accounting information.

That is one of the reasons why the MonB5G portal is extremely beneficial: it houses more than one department at a time. The MonB5G portal has a very Important database that helps contain information on things that are able to access the things that are provided by the MonB5G which is called the MonB5G subscribers database. Then there is access management which is responsible to evaluate and execute policy enforcement in regard to users being able to access the MonB5G framework. The IDMO Connector is the connector that is responsible for communicating and exchanging information with the Inter-Domain Manager and Orchestrator. Another component held within the MonB5G portal is the System Health Monitoring Component. The System Health Monitoring Component Who's responsible for containing extremely high-level monitoring in real-time and helping provide it to show currently what is happening in the network and provide it to the MonB5G System Operator. The MonB5G portal provides the slice templates with the slice tenants which help the framework.

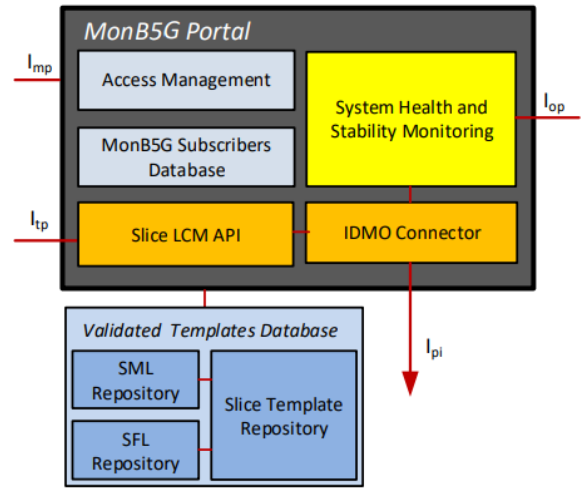

***Fig 9***. *The Above Figure Shows The MonB5G Portal Internally*

The MonB5G System Operator Has the responsibility of being able to control the stability and security of the operation of the
entire network. Because of that reason, they are allowed to be able to access all levels of the KPI and be able to execute any commands that they see fit To be able to keep all measures while doing that on the interface.

### 4.1. The Inter-Domain Manager and Orchestrator (IDMO)

The Inter-Domain Manager and Orchestrator (IDMO) is The main system which helps provides an extremely important role in preparing slices and having and deploying them helps parlay the deployment policy with the slice requester. It also helps to parlay the business dimensions to the contractor. The MonB5G interacts with the IDMO with the southbound interface which helps simplify things. The interchange of information between the two shares information like the demand for resources, the policies, and the availability of resources. This interchange of resources and information helps evaluate the circumstances that the system is in.

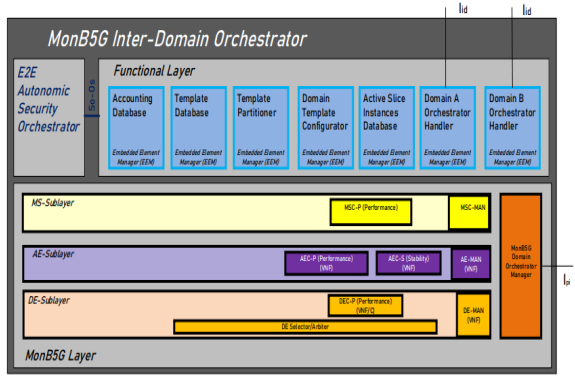

***Fig 10.*** *The Figure Above Shows The IDMO Internal Structure*

The structure of an Inter-Domain Manager and Orchestrator (IDMO) is broken down into two things the first one is the MonB5G Layers and the Functional Layers. The Domain Manager and Orchestrator (DMO) helps interact with the IDMO which helps deploy the E2E-slices. E2E Autonomic Security Orchestrator (E2E-ASO) Has the Responsibility of Maintaining security during the operation of a slice from an E2E. The partitioning of a template that will be deployed in more than one domain and the domains that are of the same type is called Template Practitioner. If there is a lack of resources in a domain that will be deployed when a Template Practitioner comes in. Template Database is similar to Template Practitioner but it helps keep all of the templates that can be used by the MonB5G framework. Then there is Template Configurator which is the process that does the first configuration of the slice. That is why each domain template that will be deployed needs to be configured by the Template Configurator. The Handler of all Orchestrators is the one who makes sure that the interactions between orchestrators that are always used by MonB5G are fully operational. This is why it is important to keep all of the history of the data and the resources that are being consumed and that is where the Accounting Database comes in.

*4.2. Next-Generation Radio Network Access (NG-RAN)*

Base stations (BSs) for LTE are autonomous from one another. The resources of the wireless spectrum are severely wasted, which is a drawback. The nearby BS interval will be disrupted by the mobile device. The cloud radio access network has been defined and selected as the next-generation RAN to address these LTE issues (NG-RAN). The NG-RAN system is made up of two linked front-haul networks, RRHs, and a pool of BBUs. With the aid of access networks, the RRHs gather wireless signals from portable devices and transfer them to the BBU pool. In order to satisfy changing traffic demands, the virtualized and centralized BBU pool may manage several base stations at once and dynamically allot spectrum, time, and space.

*4.3. Open Radio Access Network (O-RAN)*

Open Radio Access Network (O-RAN) is the next level of RAN (NG-RAN) architecture. The main goal of O-RAN was to advance upon RAN and what was the basis of 3GPP. The O-RAN Runs on the Linux Foundation and that helps execute open source software on many bases.

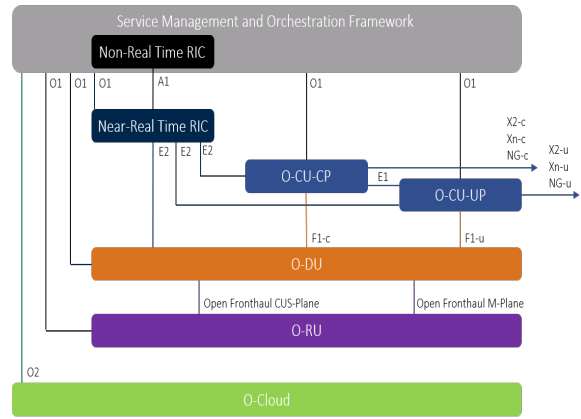

***Fig 11.*** *Service Management and Orchestration Framework*

The base unit of TS 38.401 it's called the baseband unit (BBU). The baseband unit (BBU) breaks down into two different components. The first component is called the distributed unit (DU) and the second component is called the central unit (CU). When comparing the baseband unit and the distributed unit with the central unit the distributed unit and the central unit replace the baseband unit. The two units offer what the baseband unit offers plus more things that are brand-new deployment models. These brand-new deployment models have many features like a centralized packet processing personality. This being implemented offers many different advantages even if they are or aren't sharing the same location that is located inside the 5GNodeB (gnB). The central part helps speed up the process of implementing an extremely automated process which is called multi-access edge compute (MEC) which is in coordination with RAN. Cutting that happening will also help execute the modern transport particles which will then help be in coordination with the 5G core user plane.

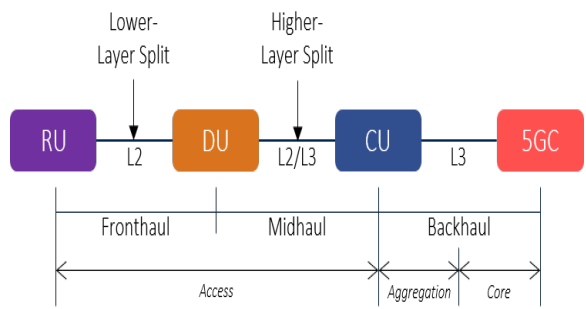

*Fig 12.* The split between the lower and higher layer

Open interfaces are either defined as a lower layer split (LLS) or a higher layer split (HLS). RU to DU is defined as a lower layer split (LLS) and DU to CU We are defined as a higher layer split (HLS). With that being established it will help portray the OSI layer in the correct way which also helps show the NG-RAN functionality. If there was a low latency service it will try to ask for a truce between the CU and DU so that way they can be partners Which will help make things a lot more cost-efficient because the CU is not located at the network core.

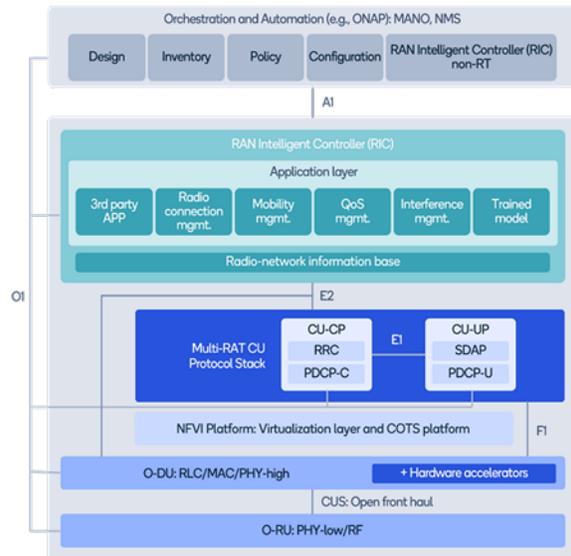

*Fig 13.* O-RAN Architecture

The above figure shows the O-RAN architecture showing the RAN Intelligent Controller (RIC) with all of the application layers and RAN base. Also presenting the Multi-RAT CU Protocol Stack. Within that, it helps show the CU-CP and CU-UP.

*4.4. eCPRI*

Enhanced Common Public Radio Interface (eCPRI) evolved from Common Public Radio Interface (CPRI) is used by many topologies for cellular networks today. Before understanding what an eCPRI is, we have to understand what CPRI is. CPRI Is a way to help translate the radio signals into a more computer-friendly way. It is an interface but hopes to move data from the RRU into the baseband. The issue arises when there is too much traffic between the baseband and the RRU. This is where eCPRI comes in to be able to help split up the traffic between the baseband and RRU making them both responsible which makes less traffic. With that, it helps open a much higher network with a lot less fiber and that will cause the mmWave to be a lot more durable and structured logically. With that being said there are only a few different cellular vendors who are taking advantage of eCPRI which is why the O-RAN This is trying to help manage the interface that the DU and RU are sharing.

## 5. CONCLUSION

In conclusion, 5G Networks Management, Orchestration, and Architecture have a major connection between the now and the future. It is a technology that will be developed throughout the near future and will become the primary technology that will be used. In this survey, we also covered some concerns regarding orchestration in management and how to be able to develop them. Development is a huge part of all 5G Network Management orchestration in architecture and the survey study that's why we spoke about AI and ML and how it will develop into something more than the task-based infrastructure that it is today.

Then we covered MonB5G architecture and its three different framework layers. The survey also covers the differences and similarities between eCPRI and CPRI. Open Radio Access Network (O-RAN) and the Inter-Domain Manager and Orchestrator were both also covered in detail to help us understand where the future of 5G Management, Orchestration, and Architecture is heading.

Future work includes extending our current cybersecurity framework [8-28] to develop a zero-trust security framework for the 5G and beyond network.